\documentclass[preprint,showpacs,preprintnumbers,
amsmath,amssymb,floatfix,pra]{revtex4}
 \usepackage{epsfig}

\usepackage{bm}

\newcommand{\be}{\begin{equation}}
\newcommand{\ee}{\end{equation}}
\newcommand{\bear}{\be\begin{array}}
\newcommand{\bea}{\begin{eqnarray}}
\newcommand{\eea}{\end{eqnarray}}
\newcommand{\nn}{\nonumber}
\newcommand{\dst}{\displaystyle}
\newcommand{\fr}[2]{\frac{{\dst #1}}{{\dst #2}}}

\date{May 31, 2006}

\begin{document}
\title{Exclusive and inclusive muon pair production in collisions of 
relativistic nuclei}
\author{K.~Hencken}
\affiliation{Institut f\"ur Physik, Universit\"at Basel, 4056, Basel,
  Switzerland}
\author{E.A.~Kuraev}
\affiliation{Joint Institute of Nuclear Reseach, 141980, Dubna, Russia}
\author{V.G.~Serbo}
\affiliation{Novosibirsk State University, 630090, Novosibirsk,
Russia}

\begin{abstract}
The exclusive production of one $\mu^+\mu^-$ pair in collisions of
two ultra-relativistic nuclei is considered. We present a simple
method for the calculation of the Born cross section for this process
based on an improved equivalent photon approximation. We find
that the Coulomb corrections to this cross section (corresponding
to multi-photon exchange of the produced $\mu^{\pm}$ with the nuclei)
are small while the unitarity corrections (corresponding to the
exchange of light-by-light blocks between nuclei) are large. This is
in sharp contrast to the exclusive $e^+e^-$ pair production where
the Coulomb corrections to the Born cross section are large while
the unitarity corrections are small. We calculate also the cross
section for the production of one $\mu^+\mu^-$ pair and several
$e^+e^-$ pairs in the leading logarithmic approximation. Using this
cross section we find that the inclusive production of $\mu^+\mu^-$
pair coincides in this approximation with its Born value.
\end{abstract}

\maketitle
\section{Introduction}

Lepton pair production in ultra-relativistic nuclear collisions
were discussed in numerous papers (see~\cite{BHTSK} for a review and
references therein). For the RHIC and LHC colliders the charge
numbers of nuclei $Z_1=Z_2\equiv Z$ and their Lorentz factors
$\gamma_1=\gamma_2\equiv \gamma$ are given in Table ~\ref{t1}.
 \begin{table}[!h]
\vspace{5mm} {\renewcommand{\arraystretch}{1.5}
 \caption{Colliders and cross sections for the lepton pair production}
\begin{center}
\par
 \begin{tabular}{|c|c|c|c|c|}\hline
Collider & $Z$ & $\gamma$ & $\sigma^{e^+e^-}_{\rm Born}$ [kb]
 & $\sigma^{\mu^+\mu^-}_{\rm Born}$ [b] \\ \hline
RHIC, Au-Au & 79 & 108 & 36.0 & 0.23
\\ \hline
LHC, Pb-Pb & 82 & 3000 & 227 & 2.6
\\ \hline
LHC, Ar-Ar & 18 & 3400 & 0.554 & 0.0082
 \\ \hline
\end{tabular}
 \label{t1}
\end{center}
 }
 \end{table}

The cross section of one $e^+e^-$ pair production in Born
approximation, described by the Feynman diagram of Fig. 1, was
obtained many years ago~\cite{Landau}. Since the Born cross section
$\sigma^{e^+e^-}_{\rm Born}$ is huge (see Table~\ref{t1}), the
$e^+e^-$  pair production can be a serious background for many
experiments. It is also an important issue for the beam lifetime and
luminosity of these colliders \cite{Klein:2000ba}. It means that
various corrections to the Born cross section, as well as, the cross
section for $n$-pair production, are of great importance. At
present, there are a lot of controversial and incorrect statements
in papers devoted to this subject. The corresponding references and
critical remarks can be found in Refs.~\cite{BHTSK,ISS-99,LMS-02}.

\begin{figure}
\begin{center}
\includegraphics[width=5cm,angle=0]{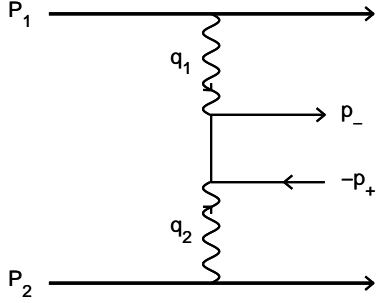}
 \caption{The Feynman diagram for the lepton pair
 production in the Born approximation}
\end{center}
 \label{F:1}
\end{figure}

Since the parameter  $Z\alpha$ may be not small ($Z\alpha \approx
0.6$ for Au-Au and Pb-Pb collisions), the whole series in
$Z\alpha$ has to be summed in order to obtain the cross section with
sufficient accuracy. The exact cross section for one pair
production $\sigma_1$ can be split into the form
 \be
\sigma_1 = \sigma_{\rm Born} + \sigma_{\rm Coul}+ \sigma_{\rm
unit}\,,
 \label{1}
 \ee
where two different types of corrections need to be distinguished.
\begin{figure}[htb]
\begin{center}
\includegraphics[width=5cm,angle=0]{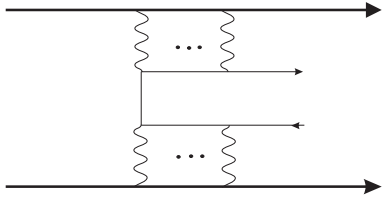}
 \caption{The Feynman diargam for the Coulomb correction}
  \label{F:2}
\end{center}
\end{figure}
The Coulomb correction $\sigma_{\rm Coul}$ corresponds to
multi-photon exchange of the produced $e^{\pm}$ with the nuclei
(Fig. \ref{F:2}); it was calculated in~\cite{ISS-99}. The unitarity
correction $\sigma_{\rm unit}$ corresponds to the exchange of
light-by-light blocks between nuclei (Fig. \ref{F:3}); it was
calculated in~\cite{LMS-02}. It was found in~\cite{ISS-99}
and~\cite{LMS-02}  that the Coulomb corrections are large while the
unitarity corrections are small (see Table~\ref{t2}). The results
of~\cite{LMS-02} were confirmed recently in~\cite{BGKN} by a direct
summation of the Feynman diagrams.
\begin{figure}[htb]
\begin{center}
\includegraphics[width=5cm,angle=0]{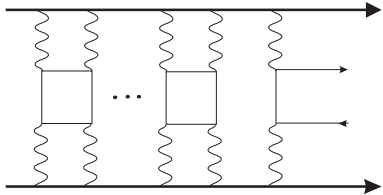}
 \caption{The Feynman diagram for the unitarity correction}
  \label{F:3}
\end{center}
\end{figure}

\begin{table}[htb]
\vspace{5mm} {\renewcommand{\arraystretch}{1.5}
 \caption{Coulomb and unitarity corrections to the $e^+e^-$ pair
 production}
\begin{center}
\par
  \begin{tabular}{|c|c|c|c|}\hline
Collider & $\frac{\displaystyle\sigma_{\rm Coul}}{\displaystyle
\sigma_{\rm Born}}$ & $\frac{\displaystyle\sigma_{\rm
unit}}{\displaystyle
\sigma_{\rm Born}}$ 
\\ \hline
RHIC, Au-Au &  $-25$\% & $-4.1$\%
\\ \hline
LHC, Pb-Pb & $-14$\% & $-3.3$\%
\\ \hline
LHC, Ar-Ar &  $-1.06$\% & $-0.025$\%
 \\ \hline
\end{tabular}
 \label{t2}
\end{center}
 }
\end{table}

In this paper we present our calculations related to the exclusive
and inclusive muon pair production. This process may be easier to
observe experimentally than $e^+e^-$ pair production described
above. It should be stressed that the calculation scheme, as well as,
the final results for the $\mu^+\mu^-$ pair production are quite
different than those for the $e^+e^-$ pair production.

In the next section we calculate the Born cross section for one
$\mu^+\mu^-$ pair production using the improved equivalent photon
approximation with an accuracy about 5 \%. In Sect. 3 we present the
Coulomb and unitarity corrections to the exclusive production of one
muon pair. In this section we also obtain the probability for the
production of one muon and several electron pairs in collisions of
nuclei at a given impact parameter $\rho$. This result allows us to
calculate the inclusive production of one muon pair in Sect. 4. In
Sect. 5 we calculate the cross section for two muon pair production
in lowest order. The conclusion of the paper is given in Sect.
6. In the Appendix we consider the relatively simple process of one muon
pair production by a real photon off a nucleus $\gamma Z \to Z
\mu^+\mu^-$ which serves as a good test of the approach used. A
short preliminary version  of this paper was published
in~\cite{KKS-2005}.

\section{Born cross section for one $\mu^+\mu^-$ pair
production}

The production of one {$\mu^{+}\mu^{-}$} pair
  \be
Z_1+Z_2 \to Z_1+Z_2+ \mu^+\mu^-
 \label{2}
 \ee
in the Born approximation is described by the Feynman diagram of
Fig. 1. When two nuclei with charges $Z_1e$ and $Z_2e$ and 4-momenta
$P_1$ and $P_2$ collide with each other, they emit equivalent
(virtual) photons with the 4--momenta $q_1$, $q_2$, energies
$\omega_1$, $\omega_2$ and their virtualities $Q_1^2=-q_1^2$,
$Q_2^2=-q_2^2$. Upon fusion, these photons produce a
{$\mu^{+}\mu^{-}$} pair with the total four--momentum $q_1+q_2$ and
the invariant mass squared $W^2 = (q_1 +q_2)^2$. Besides this we denote
 $$
(P_1+P_2)^2=4E^2=4M^2\,\gamma^2\,,\;\; \alpha \approx 1/137
 $$
and use the system of units in which $c=1$ and $\hbar =1$.

\subsection{General formulae}

The Born cross section of the process (\ref{2}) can be calculated
to a good accuracy using the equivalent photon approximation (EPA)
in the improved variant presented, for example, in Ref.~\cite{BGMS}.
Let the numbers of equivalent photons be $dn_1$ and $dn_2$. The most
important contribution to the production cross section stems from
photons with very small virtualities $Q_i^2 \ll {\mu}^2$ where $\mu$
is the muon mass. Therefore to a good approximation, the photons move in
opposite directions, and $W^2 \approx 4\,\omega_1\,
\omega_2$. In this very region the Born differential cross section
$d\sigma_{\rm B}$ for the process considered is related to the cross
section  $\sigma_{\gamma\gamma}$ for the process with real photons: $\gamma\gamma \to
\mu^{+}\mu^{-}$ by the equation
\begin{equation}
d\sigma_{\rm B} = dn_1 dn_2 \, d\sigma_{\gamma \gamma} (W^2)\,,
\;\;\; W^2 \approx 4\,\omega_1\,\omega_2\,.
 \label{3}
\end{equation}
The number of equivalent photons are (see Eq.~($D.4$) in
Ref.~\cite{BGMS})
 \begin{equation}
dn_i(\omega_i,Q_i^2) = \frac{Z_i^2 \alpha}{\pi}\,\left(1-\frac{\omega_i
}{E_i} \right)\, \frac{d\omega_i}{\omega_i}\, 
\left( 1- \frac{Q^2_{i\,\min}}{Q_i^2} \right)
\,F^2(Q^2_i) \, \frac{dQ^2_i}{Q^2_i} \,,
 \label{4}
\end{equation}
where
 \be
Q_i^2\ge Q^2_{i\,\min} = \frac{\omega_i^2}{\gamma^2}
 \label{5}
 \ee
and $F(Q^2)$ is the electromagnetic form factor of the nucleus. It is
important that the integral over $Q^2$ converges rapidly for $Q^2
> 1/R^2$, where
 \be
R=1.2\,A^{1/3}\;\; {\rm fm}
 \label{6}
 \ee
is the radius of the nucleus with $A\approx M/m_p$ the number of
nucleons ($R\approx 7$ fm, $1/R\approx 28$ MeV for Au and Pb). Since
$Q^2_{\min} \lesssim 1/R^2$, the main contribution to the cross
section is given by virtual photons with energies
\begin{equation}
\omega_i \lesssim \gamma/R\,.
 \label{7}
\end{equation}
Therefore, we can use the spectrum of equivalent photons
neglecting terms proportional to $\omega_i /E_i$ given by:
\begin{equation}
dn_i(\omega_i,Q_i^2) = \frac{Z_i^2 \alpha }{ \pi}\, \frac{d\omega_i }{
 \omega_i}\, \left( 1- \frac{\omega_i^2}{\gamma^2 Q_i^2 } \right)
\,F^2(Q^2_i) \, \frac{dQ^2_i}{Q^2_i} \,.
 \label{8}
\end{equation}

After the transformation
\begin{equation}
\frac{d\omega_1}{\omega_1}\; \frac{d\omega_2}{ \omega_2}\; =
\frac{d\omega_1 }{ \omega_1} \, \frac{dW^2 }{ W^2}
 \label{9}
\end{equation}
we cast the cross section in the form
 \bea
  d\sigma_{\rm B}&=& \frac{Z_1^2\,Z_2^2 \alpha^2}{
\pi^2 }\, \frac{ d\omega_1}{ \omega_1}\,\left( 1- \frac{\omega_1^2}{
\gamma^2 Q_1^2 }\right) \,F^2(Q^2_1) \, \frac{dQ^2_1}{ Q^2_1}\times
   \label{10}\\
&&\left( 1- \frac{\omega_2^2}{\gamma^2 Q_2^2 } \right) \,F^2(Q^2_2)
\, \frac{dQ^2_2}{Q^2_2} \, \frac{dW^2}{W^2}\,\sigma_{\gamma \gamma}
(W^2) \,,
 \nn
 \eea
where $\omega_2\approx W^2/(4 \omega_1$).

\subsection{Leading logarithmic approximation (LLA)}

Before using the calculation scheme above, it is instructive to
present a rougher but simpler approximation --- the so called
leading logarithmic approximation (LLA). In the LLA, the equivalent
photon spectrum as a function of photon energy $dn_i (\omega_i)$ is obtained after
integrating $dn_i(\omega_i ,Q_i^2)$ over $Q_i^2$ in the region between
 \be
Q^2_{i\,\min}\le Q_i^2 \lesssim 1/R^2
 \label{11}
  \ee
which leads to
 \be
dn_i(\omega_i) \approx \frac{Z_i^2 \alpha }{\pi} \,
\ln\frac{\gamma^2}{ (R\omega_i)^2}\, \frac{d\omega_i}{\omega_i}\,.
 \label{12}
\end{equation}
The restriction $Q^2_{i\,\min} \lesssim 1/ R^2$ corresponds to the
integration interval
 \be
a=\frac{W^2 R}{ 4 \gamma} \lesssim \omega_1 \lesssim b=
 \frac{\gamma}{ R}\,,
 \label{13}
  \ee
which gives
\begin{equation}
 \label{14}
\sigma^{{\rm LLA}}_{\rm B}=\frac{Z_1^2 Z_2^2 \alpha^2 }{ \pi^2 }\,
\int_{4\mu^2}^\infty\,\frac{dW^2 }{ W^2}\,\sigma_{\gamma \gamma}
(W^2)\,\int_a^b\frac{d\omega_1}{ \omega_1}\,\ln\frac{b^2}{
\omega_1^2}\, \ln\frac{\omega^2_1}{ a^2}\,.
\end{equation}
Since $\sigma_{\gamma \gamma} (W^2)\approx (4\pi \alpha^2/W^2)\,
\ln{(W^2/\mu^2)}$ for large values of $W\gg \mu$, the main
contribution to the Born cross section comes from the region of
small values of $W$ near the threshold. Therefore, within
logarithmic accuracy we replace $W$ by some fixed value $W_0\sim
2\mu$ in the lower bound $a$. After that the integral over $W^2$ gives
\be
I=\int_{4\mu^2}^\infty \frac{dW^2 }{ W^2} \, \sigma_{\gamma \gamma}
(W^2) =\frac{14\pi \alpha^2}{ 9\mu^2}
 \label{15}
 \ee
and further integration over $\omega_1$ leads to
 \be
\int_a^b\frac{d\omega_1}{ \omega_1}\,\ln\frac{b^2}{ \omega_1^2}\,
\ln\frac{\omega^2_1}{ a^2}=\frac{2}{ 3} \,L^3\,,
 \label{16}
 \ee
where
 \be
L= \ln\frac{\gamma^2}{ (W_0 R/2)^2}\,.
 \label{17}
 \ee

As a result, we obtain
\begin{equation}
 \label{18}
\sigma^{{\rm LLA}}_{\rm B} =\frac{28}{ 27 \pi} \frac{(Z_1\alpha
Z_2\alpha)^2  }{ \mu^2} \, L^3
\end{equation}
in accordance with the result of Landau \& Lifshitz~\cite{Landau}. The
accuracy of the LLA  depends on the choice of the value for $W_0$. If we
use for numerical estimations $W_0=3\mu$, then the accuracy of the
LLA for the colliders discussed is about 15\%.

The same result can be obtained in the framework of the impact
parameter dependent representation, which will also be useful later.
For this aim we introduce the probability for muon pair production
$P_{\rm B}(\rho)$ in the collision of two nuclei at a fixed impact
parameter $\rho$. For $\gamma \gg 1$, it is possible to treat the
nuclei as sources of external fields and to calculate $P_{\rm
B}(\rho)$ analytically using the same approach as in
Ref.~\cite{LMS-02}. The Born cross section $\sigma_{\rm B}$ can then
be obtained by the integration of $P_{\rm B}(\rho)$ over the impact
parameter:
 \be
  \label{19}
 \sigma_{\rm B}=\int \,P_{\rm B}(\rho)\,d^2{\rho}\,.
 \ee
We calculate this probability in the LLA using Eq.~(\ref{3}) with
\begin{equation}
 dn_i = \frac{Z_i^2\alpha}{ \pi^2}\, \frac{d\omega_i }{ \omega_i}\, \frac{d^2
\rho_i}{ \rho_i^2}\,; \; \quad\omega_i \ll \frac{\gamma}{R}\,;
\;\; R\ll \rho_i \ll \frac{\gamma}{\omega_i}\,,
 \label{20}
\end{equation}
where $\bm\rho_i$ is the impact parameter of $i$-th equivalent
photon with respect to the $i$-th nucleus. This allows us to write
the above probability in the form
 \begin{equation}
P_{\rm B}(\rho) = \int dn_1 dn_2 \,\delta(\bm \rho_1 -\bm \rho_2-
\bm \rho)\,\sigma_{\gamma\gamma}(W^2)=
\frac{28}{9\pi^2}\,\frac{\left(Z_1 \alpha
Z_2\alpha\right)^2}{(\mu\rho)^2}\, \Phi(\rho)\,.
 \label{21}
\end{equation}
Depending on the value of $\rho$ two different forms for $\Phi(\rho)$
need to be used:
 \bea
\Phi(\rho)=& \left(4\ln{ \fr{\gamma}{\mu\rho}}
+\ln{\fr{\rho}{R}}\right)\, \ln{\fr{\rho}{R}}\; \;\;\;\mbox{for}\;\;&
R \ll \rho \le \fr{\gamma}{\mu}\,,
 \label{22}\\
\Phi(\rho)=&  \left(\ln{\fr{\gamma^2}{\mu^2  \rho R}}\right)^2
\;\;\;\;\;\;\;\;\;\;\;\;\;\;\;\;\;\;\;\mbox{for}\;\;&\fr{\gamma}{\mu}
\le \rho \ll \fr{\gamma^2}{\mu^2 R}\,.
  \label{23}
 \eea
Note that the function $\Phi(\rho)$ is continuous at
$\rho= \gamma/\mu$ together with its first derivative.
As expected the integration of $P_{\rm B}(\rho)$ over $\bm \rho$ in
the region $R < \rho < \gamma^2/(\mu^2 R)$ gives back the result in
(\ref{18}).

To prove (\ref{21})--(\ref{23}), we make the transformation given
in~(\ref{9}) together with the integration over $W^2$ according
to~(\ref{15}). This gives
\begin{eqnarray}
P_{\rm B}(\rho)=\frac{14}{9\pi^3}\frac{(Z_1\alpha Z_2\alpha)^2}
{\mu^2} \int_{\mu^2 R/\gamma}^{\gamma/R}\frac{d\omega_1}{\omega_1}
\int\frac{d^2\rho_1}{\rho_1^2(\bm \rho-\bm\rho_1)^2}\,
\vartheta\left(\frac{\gamma}{\omega_1}-\rho_1\right)
\vartheta\left(\frac{\gamma\omega_1}{\mu^2}-|\bm\rho-\bm
\rho_1|\right)
 \label{24}
\end{eqnarray}
where $\vartheta(x)$ is the step function. The main contribution to
this integral is given by two regions: $R\ll \rho_1\ll \rho$ and
$R\ll |\bm\rho-\bm \rho_1|=\rho_2\ll \rho$. Moreover, the two regions in
the $\omega_1$ integration with $\mu < \omega_1 < \gamma/R$ and $\mu^2
R/ \gamma < \omega_1 < \mu$ give the same contributions. As a
result we get
 \bea
\Phi(\rho)&=& 2\left(J_1+J_2\right)\,,
 \label{25}\\
J_1&=& \int_{\mu}^{\gamma/R}\frac{d\omega_1}{\omega_1}
\int_{R}^\rho\frac{d\rho_1}{\rho_1}\,
\vartheta(\gamma/\omega_1-\rho_1)\,
\vartheta(\gamma\omega_1/\mu^2-\rho)\,,
 \nn\\
J_2&=& \int_{\mu}^{\gamma/R}\frac{d\omega_1}{\omega_1}
\int_{R}^\rho\frac{d\rho_2}{\rho_2}\,\vartheta(\gamma/\omega_1-\rho)\,
\vartheta(\gamma\omega_1/\mu^2-\rho_2)\, .
 \nn
\end{eqnarray}
Next we consider the two regions of $\rho$.

In the region of relatively small impact parameters, $R \ll \rho \le
\gamma/\mu$, the second step function does not impose any
limitations, therefore,
 \bea
J_1&=&\int_{\mu}^{\gamma/\rho}\frac{d\omega_1}{\omega_1}
\int_{R}^\rho\frac{d\rho_1}{\rho_1}
+\int_{\gamma/\rho}^{\gamma/R}\frac{d\omega_1}{\omega_1}
\int_{R}^{\gamma/\omega_1}\frac{d\rho_1}{\rho_1}= \ln\frac{ \gamma
}{ \mu \rho}\,\ln\frac{\rho }{ R}+ \frac{1}{ 2}
 \left(\ln\frac{\rho}{ R}\right)^2
\nn
 \\
J_2&=&\int_{\mu}^{\gamma/\rho}\frac{d\omega_1}{\omega_1}
\int_{R}^\rho\frac{d\rho_1}{\rho_1}
 = \ln\frac{ \gamma }{ \mu \rho}\,\ln\frac{\rho }{ R}\, .
  \nn
  \eea
Summing up, we obtain (\ref{22})

In the region of relatively large impact parameters, $\gamma/\mu
\le \rho \ll \gamma^2/(\mu^2 R)$, we have
 \bea
 J_1=\int_{\mu^2  \rho/\gamma}^{\gamma/R}\frac{d\omega_1}{\omega_1}
\int_{R}^{\gamma/\omega_1}\frac{d\rho_1}{\rho_1}
 = \frac{1}{ 2} \left(\ln\frac{ \gamma^2 }{ \mu^2 R
 \rho}\right)^2\,,\;\; J_2=0\,,
  \nn
  \eea
therefore, the sum gives (\ref{23}).

We compare Eqs.~(\ref{21})--(\ref{23}) for $\Phi(\rho)$ with the
numerical calculations based on the exact matrix element calculated
with the approach as outlined in \cite{HTB1994}. We find good
agreement for Pb-Pb collisions: the discrepancy is less then 10~\%
at $\mu \rho > 10$ and it is less then 15~\%  at $\mu \rho > 2\mu R
=7.55$.

\subsection{More refined calculation}

In the calculation below we use for the form factor of the nucleus
the simple approximation of a monopole form factor, which
corresponds to an exponentially decreasing charge distribution,
whose mean squared radius $\sqrt{\left\langle r^2 \right\rangle}$ is
adjusted to the experimental value:
\begin{equation}
 \label{26}
F(Q^2) = \frac{1}{1+Q^2/\Lambda^2}\,,\;\;\; \Lambda^2 = \frac{6 }{
\left\langle r^2 \right\rangle}.
\end{equation}
For lead and gold, the parameter is $\Lambda \approx 80\,{\rm MeV}$.
This approximate form of the form factor enables us to perform some
calculations analytically, which otherwise could only be done
numerically.

\begin{figure}[htb]
\begin{center}
\includegraphics[width=9cm,angle=0]{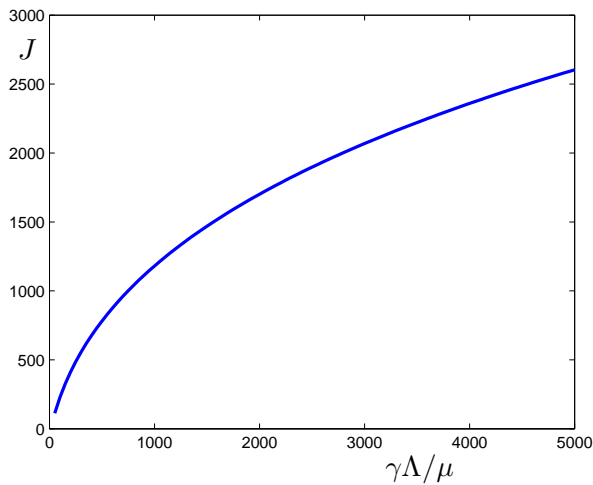}
 \caption{The function $J(\gamma\Lambda/\mu)$ from Eq. (\ref{31})}
  \label{F:4}
\end{center}
\end{figure}

The equivalent photon spectrum $dn_i (\omega_i)$ is obtained after
integrating $dn_i(\omega_i ,Q_i^2)$ over $Q_i^2$ (the upper limit
of this integration can be set to be equal to infinity in a good
approximation, due to the fast convergence of the integral at
$Q^2>\Lambda^2$):
\begin{equation}
 \label{27}
dn_i(\omega_i) = \frac{Z_i^2 \alpha }{ \pi} \, f\left(\frac{\omega_i
}{ \Lambda \gamma}\right)\, \frac{d\omega_i}{ \omega_i}\,.
\end{equation}
Here the function
\begin{equation}
 \label{28}
f(x)= (1 + 2\,x^2)\, \ln{\left(\frac{1}{ x^2} +1\right)}\, -\, 2
\end{equation}
is large for small values of $x$,
 \be
f(x)\approx \ln\frac{1}{ x^2}\, -\, 2= \ln\frac{1}{ ({\rm e}x)^2} \;\; {\rm at}
\;\; x\ll 1 \,,
 \label{29}
 \ee
but drops very quickly for large $x$ in accordance with
Eq.~(\ref{7}):
 \be
f(x) < \frac{1}{ 6\,x^4} \;\; {\rm  for}\;\; x > 1 \,.
 \label{30}
 \ee
Finally we obtain
 \be
\sigma_{\rm B} =\frac{Z_1^2 Z_2^2 \alpha^2}{ \pi^2 }\,
\int_{4\mu^2}^\infty\,\frac{dW^2 }{ W^2}\, G(W^2)\,\sigma_{\gamma
\gamma} (W^2)= \frac{\left(Z_1 \alpha Z_2\alpha\right)^2 }{ \pi
\mu^2}\, J(\gamma \Lambda/\mu)\,,
 \label{31}
\end{equation}
where
 \be
G(W^2)= \int\limits^{\omega_{\max}}_{\omega_{\min}} \frac{d\omega }{
\omega}\; f\left(\frac{\omega }{ \Lambda \gamma}\right)\,
f\left(\frac{W^2 }{ 4\Lambda \gamma\omega}\right)\,.
 \label{32}
  \ee
Since $\omega_i < E$  and $\omega_1 \omega_2 \sim  \mu^2$ we have
$\omega_{\min} \sim\mu^2/E$ and $\omega_{\max} =E$. However, due to
the fast decrease of $f(x)$ for $x>1$ one can extend these limits up to
$\omega_{\min}= 0$ and $\omega_{\max} = \infty$ without any lack of
accuracy, therefore,
 \be
G(W^2)= 2\,\int\limits^{\infty}_{0}  f(x_1)\, f(x_2)\, dy\,,\;\;
 x_{1,2} = \frac{W}{ 2\Lambda \gamma}\, {\rm e}^{\pm y}\,.
  \label{33}
  \ee
A numerical evaluation of the integrals in Eqs.~(\ref{31}),
(\ref{32}) yields the function $J(\gamma\Lambda/\mu)$ presented in
Fig. \ref{F:4}.

Corrections to the photon spectrum are represented by terms in
Eqs.~(\ref{3}), (\ref{8}) of the order of $Q^2_i/W^2 $
(see Eqs. (E.1) in Ref.~\cite{BGMS}), which are dropped before
the integration over $Q_i^2$ is done. After the integration with weight
$1/Q_i^2$ the relative value of these corrections becomes of the
order of
\begin{equation}
\eta_1 = \frac{\Lambda^2 }{ W^2 \, L}\,.
 \label{34}
\end{equation}
Thus, for the collisions considered here one can estimate the accuracy of
the calculations on the level $\eta_1\sim 5 \% $. Another test of
accuracy of the approach used is given in Appendix.

Note that in the LLA the function $G(W^2)$ is just
\begin{equation}
G^{{\rm LLA}}(W^2) = \frac{2}{ 3} \left[\ln\frac{\gamma^2}{ ({\rm
e}W/2\Lambda)^2} \right]^3\approx \frac{2}{ 3} \left[\ln\frac{\gamma^2}{
(WR/2)^2} \right]^3
 \label{35}
\end{equation}
in accordance with Eq.~(\ref{16}) (taking into account that
$\Lambda/{\rm e} \approx 1/R$).

\section{Coulomb and unitarity corrections}

The Coulomb correction corresponds to Feynman diagram of Fig. 2. Due
to the restriction of the transverse momenta of additionally
exchanged photons to the range below $1/R$, the effective parameter
of the perturbation series is not $(Z\alpha)^2$ but
$(Z\alpha)^2/(R\mu)^2$. In addition, the contribution of the
additional photons is suppressed by a logarithmic factor. Indeed,
the cross section for two--photon production mechanism is
proportional to $L^3$, while the cross section for the
multiple-photon production mechanism is proportional only to $L^2$.
Therefore, the real parameter describing the suppression of the
Coulomb correction is of the order of
 \be
\eta_2= \frac{(Z\alpha)^2 }{ (R \mu )^2 L}
 \label{36}
  \ee
which corresponds to Coulomb corrections of less then $1\,$\%. The
example considered in Appendix confirms this estimate.

The unitarity correction $\sigma_{\rm unit}$ to the one muon pair
production corresponds to the exchange of light-by-light blocks
between the two nuclei (Fig. 3). We start with a more general process ---
the production of one $\mu^+\mu^-$ pair and $n$ electron-positron
pairs ($n\geq 0$) in a collision of two ultra-relativistic nuclei
 \be
Z_1+Z_2 \to Z_1+Z_2+\mu^+\mu^- +n\, (e^+e^-)
 \label{37}
 \ee
taking into account the unitarity corrections, which correspond to
the exchange of the blocks of light-by-light scattering via the
virtual lepton loops. The corresponding cross section
$d\sigma_{1+n}$ can be calculated by a simple generalization of the
results obtained in~\cite{BGKN} for the $n$-pair process without
muon pair production: $Z_1+Z_2 \to Z_1+Z_2+n\, (e^+e^-)$. This
multiple pair production process was studied in a number of papers,
see \cite{BHTSK} for a review. It was found that the probability is
to a good approximation given by a Poisson distribution with the
deviations found to be small. Indeed, it is not difficult to show
that the basic equations for the latter process should be modified
as follows. In Eq. (26) of~\cite{BGKN} the additional factor
 \be
\tilde{B}^\mu(\bm\rho, {\bf r}_{n+1})\,{\rm
e}^{-L[A^\mu(\rho)/2+{\rm i} \varphi^\mu(\rho)]}
 \label{38}
 \ee
appears under the integral, where $L=\ln{(\gamma_1 \gamma_2)}$
and the functions $\tilde{B}^\mu$, $A^\mu$ and $\varphi^\mu$ are
the same as the functions $\tilde{B}$, $A$ and $\varphi$ in Eq.
(27) of~\cite{BGKN} but with the replacement of electrons by
muons. As a result, Eq.~(31) of~\cite{BGKN} is replaced by
 \be
\frac{d\sigma_{1+n}}{ d^2\rho} = LA_1^\mu(\rho)\,
\frac{[LA_1(\rho)]^n}{ n!}\,{\rm e}^{-LA_1^\mu(\rho)-LA_1(\rho)}\,,
 \label{39}
 \ee
where $LA_1^\mu(\rho)\approx P_{\rm B} (\rho)$ is the probability
for one muon pair production in the Born approximation, as discussed in
Sect. 2.2. In the region of interest, $\rho > 2R$, the function
$A_1^\mu(\rho)$ is small,
 \be
LA_1^\mu(\rho) \ll 1\,,\;\; A_1^\mu(\rho)\ll A_1(\rho)\,,
 \label{40}
 \ee
therefore,  we can rewrite (\ref{39}) in the simpler form
  \be
\frac{d\sigma_{1+n}}{ d^2\rho} =P_{1+n} (\rho)\,,\;\; P_{1+n}
(\rho)=P_{\rm B} (\rho) \, \frac{[\bar{n}_e(\rho)]^n}{ n!}\,{\rm
e}^{-\bar{n}_e(\rho)}\,,
 \label{41}
 \ee
where $\bar{n}_e(\rho)=LA_1(\rho)$ is the average number of $e^+e^-$
pairs produced in collisions of the two nuclei at a given impact
parameter $\rho$. The result that the probabilities for the
different processes factories is due to the independence of the
individual processes. For a general discussion of the validity of
this factorization together with possible violations we refer to
\cite{Baur:2003ar}.

In particular, we get the cross section for the exclusive one
$\mu^+\mu^-$ pair production including the unitarity correction as
 \be
\sigma_{1+0} = \int P_{\rm B} (\rho)\,{\rm e}^{-\bar{n}_e(\rho)}
\,d^2\rho\,.
 \label{42}
 \ee
This expression can be rewritten in the form
 \be
\sigma_{1+0} = \sigma_{\rm B} +\sigma_{\rm unit}\,,
 \label{43}
 \ee
where
 \be
\sigma_{\rm B}=\int P_{\rm B} (\rho)\,\,d^2\rho
 \label{44}
 \ee
is the Born cross section discussed in Sect. 2 and
 \be
\sigma_{\rm unit}=- \int \left[1-{\rm e}^{-\bar{n}_e(\rho)}\right]\,
P_{\rm B} (\rho)\,d^2\rho
 \label{45}
 \ee
corresponds to the unitarity correction for the one muon pair
production.

A rough estimation of $\sigma_{\rm unit}$ can be done
as follows. The main contribution to $\sigma_{\rm unit}$ comes
from the region
 \be
 \label{46}
R \ll \rho \ll 1/m_e
  \ee
in which the function $\bar{n}_e(\rho)\approx \bar{n}_e(2R)$ and
the integral (\ref{45}) can be calculated in LLA. It gives
 \be
 \label{47}
\sigma_{\rm unit}\sim -\frac{28}{ 27 \pi} \frac{(Z_1\alpha Z_2\alpha)^2
}{ \mu^2} \,\left[1-\mbox{e}^{- \bar{n}_e(2R)}\right] \,J_{\rm
unit}\,,
 \ee
where
 \be
 \label{48}
  J_{\rm unit} = 6\int_{2R}^{1/m_e}\, \Phi(\rho) \,\frac{d\rho}{ \rho}\,.
  \ee
As a result, we find $\sigma_{\rm unit}\sim -1.2$ barn for the Pb-Pb
collisions at LHC, which corresponds approximately to $(-50)$\% of
the Born cross section.

It is seen that unitarity corrections are large, in other words, the
exclusive production of one muon pair differs considerable from its
Born value.

\section{Inclusive production of one $\mu^+\mu^-$ pair}

The experimental study of the exclusive muon pair production seems
to be a very difficult task. Indeed, this process requires that the
muon pair should be registered without any electron-positron pair
production including $e^\pm$ emitted at very small angles.
Otherwise, the corresponding cross section will be close to the Born
cross section.

To prove this, let us consider the process (\ref{37}), whose
probability is given by Eq. (\ref{41}). The corresponding cross
section is
 \be
 \sigma_{1+n} = \int\,P_{1+n}(\rho)\, d^2\rho\,.
  \label{49}
 \ee
It is clearly seen from this equation that after summing up over all
possible electron pairs we obtain the Born cross section
 \be
 \sum_{n=0}^{\infty}\, \sigma_{1+n} = \sigma_{\rm B}\,.
   \label{50}
 \ee
Therefore, there is a very definite prediction: the inclusive muon
pair production coincides with the Born limit. This direct
consequence of calculations, which take into account strong field
effects, may be easier to test experimentally than the prediction
for cross sections of several $e^+e^-$ pair production.

\section{Two muon pair production}

The cross section of the process
 \be
Z_1 Z_2 \to Z_1 Z_2 + \mu^+\mu^-\mu^+\mu^-
  \label{51}
 \ee
can be calculated in lowest order in $\alpha$ according to
 \be
\sigma_2 = \frac{1}{ 2} \int \left[P_{\rm B}(\rho)\right]^2 d^2\rho
  \label{52}
 \ee
with the integration region $\rho \geq 2R$. But in this region the
probability $P_{\rm B}(\rho)$ is given to a good accuracy by
Eqs.~(\ref{23})--(\ref{25}). From this we get $\sigma_{2} = 1.24$
mbarn for Pb-Pb collisions at LHC.

 \section{Conclusion}

The exclusive production of one $\mu^+\mu^-$ pair in collisions of
two ultra-relativistic nuclei is considered. We present a simple
method for the calculation of the Born cross section for this process.

We found that the Coulomb corrections to this cross section
are small, while the unitarity corrections are large. This is in
sharp contrast to the exclusive $e^+e^-$ pair production where the
Coulomb corrections to the Born cross section are large while the
unitarity corrections are small.

We calculate also the cross section for the production of one
$\mu^+\mu^-$ pair and several $e^+e^-$ pairs in LLA. Using this
cross section we found that the inclusive production of $\mu^+\mu^-$
pair coincides in this approximation with its Born value.

Let us discuss the relation of the cross sections obtained for the
muon pair production with the the differential cross section of the
$e^+e^-$ pair production in the region of large transverse momenta
for the $e^{\pm}$, for example at $p_{\pm\, \perp} \gtrsim 100$ MeV. It
is clear that for the $e^+e^-$ pair production in this region, the
situation is similar to the case considered for $\mu^+\mu^-$ pair
production.

We expect that the inclusive production of a single $e^+e^-$ pair
with large transverse momenta of $e^{\pm}$ (together with several
unobserved $e^+e^-$ pairs in the region of small transverse momenta
of $e^{\pm}$ of the order of $m_e$) coincides with the Born limit.

 \section*{Appendix}

To tests the approach used in Sect. 2.3, we consider the simpler
case of the muon pair production by a real photon with the
energy $\omega$ off an nucleus
 \be
\gamma Z \to Z \mu^+\mu^-\,.
  \label{A1}
 \ee
This cross section was calculated by Ivanov and Melnikov in
Ref.~\cite{IM98} using the same expression (\ref{26}) for the form
factor of the nucleus and assuming $\Lambda^2/(2\mu)^2 \ll 1$. The
corresponding formula for the Born contribution and the first
Coulomb correction is
 \be
\sigma_{\gamma Z} = \frac{28}{ 9}\,\frac{Z^2 \alpha^3}{
\mu^2}\,\left(l-C_1-C_2\right) \,,
  \label{A2}
   \ee
where
 \be
l= \ln\frac{2\omega \Lambda}{ \mu^2}- \frac{57}{ 14}\,,\;\; C_1=
\frac{12}{ 35} \left(\frac{\Lambda}{ 2\mu}\right)^2\,,\;\; C_2= 0.92\,
(Z\alpha)^2\, C_1\,.
  \label{A3}
 \ee
Therefore, the relative magnitude of the Coulomb correction is given by
 \be
\eta_2= \frac{C_2}{ l}\,,
 \label{A4}
 \ee
which confirms the estimate in (\ref{36}).

In the equivalent photon approximation, the cross section is given by
 \be
d\sigma_{\gamma Z}^{\rm EPA}= dn_2 \,\sigma_{\gamma \gamma}(W^2)
 \label{A5}
  \ee
which has the form
 \be
\sigma_{\gamma Z}^{\rm EPA}=\frac{Z^2 \alpha}{ \pi}
\int_{4\mu^2}^\infty\,\frac{dW^2 }{ W^2}\, f\left(\frac{W^2}{ 2\omega
\Lambda}\right)\,\sigma_{\gamma \gamma} (W^2)\,.
 \label{A6}
  \ee
The main contribution to this integral is given by the region near
the lower limit, where the argument of the function $f$ is small
and therefore $f$ can be replaced by its approximate expression
(\ref{29}):
 \be
f\left(\frac{W^2}{ 2\omega \Lambda}\right)= 2\,\ln\frac{\omega \Lambda}{
2 \mu^2}-2  -2\, \ln\frac{W^2}{ 4\mu^2}\,.
 \label{A7}
 \ee
After that the cross section can be calculated without difficulties as
 \be
\sigma_{\gamma Z}^{\rm EPA}=2\,\frac{Z^2 \alpha}{ \pi}\left[
\left(\ln\frac{\omega \Lambda}{ 2 \mu^2}-1\right)\,I- I_1\right]=
\frac{28}{ 9}\,\frac{Z^2 \alpha^3}{ \mu^2}\;l\,,
 \label{A8}
  \ee
where $I$ is given by (\ref{15}) and
 \be
I_1=\int_{4\mu^2}^\infty \frac{dW^2 }{ W^2} \, \sigma_{\gamma
\gamma} (W^2) \, \ln\frac{W^2}{ 4\mu^2} =\frac{(43-28\ln{2})\,
\pi\alpha^2}{ 9\mu^2}\,.
 \label{A9}
 \ee
Comparing this expression with the one of (\ref{A2}), we find that
those terms, which are omitted in the EPA, have a relative magnitude
of the order of
 \be
\eta_1= \frac{C_1}{ l}\,,
 \label{A10}
 \ee
this expression confirms the estimate (\ref{34}).

\section*{Acknowledgments}

We are grateful to D.~Ivanov and A.~Milstein for useful discussions.
This work is supported by RFBR (code 06-02-16064).


\end{document}